\documentclass[aps,pra,reprint,groupedaddress]{revtex4-1}
\usepackage{graphicx}
\usepackage{amssymb}
\usepackage{amsmath}
\usepackage{xfrac}

\usepackage{color}
\usepackage{soul}

\definecolor{orange}{rgb}{1,0.7,0.4}
\definecolor{red}{rgb}{1,0.3,0.3}

\begin{document}

\title{Excitons in asymmetric quantum wells}

\author{P. S. Grigoryev, A. S. Kurdyubov, M. S. Kuznetsova, Yu. P. Efimov, S. A. Eliseev, V. V. Petrov, V. A. Lovtcius, P. Yu. Shapochkin, and I. V. Ignatiev}

\affiliation{Spin optics laboratory, SPbU}
\affiliation{Nanophotonics, SPbU}

\date{\today}

\begin{abstract}
Resonance dielectric response of excitons is studied for the high-quality GaAs/InGaAs heterostructures with wide asymmetric quantum wells (QWs). To highlight effects of the QW asymmetry, we have grown and studied several heterostructures with nominally square QWs as well as with  triangle-like QWs. Several quantum confined exciton states are experimentally observed as narrow exciton resonances with various profiles. A standard approach for the phenomenological analysis of the profiles is generalized by introducing of different phase shifts for the light waves reflected from the QWs at different exciton resonances. Perfect agreement of the phenomenological fit to the experimentally observed exciton spectra for high-quality structures allowed us to obtain reliable parameters of the exciton resonances including the exciton transition energies, the radiative broadenings, and the phase shifts. A direct numerical solution of Schr\"{o}dinger equation for the heavy-hole excitons in asymmetric QWs is used for microscopic modeling of the exciton resonances.
Remarkable agreement with the experiment is achieved when the effect of indium segregation during the heterostructure growth is taken into account. The segregation results in a modification of the potential profile, in particular, in an asymmetry of the nominally square QWs.    
\end{abstract}

\pacs{}

\maketitle

\section{Introduction}
Excitons in quantum wells (QWs) are extensively studied already for four decades~\cite{Dingle-PRL1974, Sham-PRL1981, Bastard-book, Davies-book, Ivchenko-book, Kavokin-book}.  Theoretical analysis of the excitons typically assumes a simplified model for the QW potential, e.g., square profiles for electrons and holes~\cite{Ivchenko-book, Andreani-SSC1991, Ivchenko-FTT1991, Tassone-PRB1992}. Real potential, however, is more complex due to several processes occurring during the growth of heterostructures. In the narrow QWs, the monolayer fluctuations of interfaces give rise to fluctuations of the QW width. These fluctuations result in the step-like changes of the exciton quantization energy experimentally observed as a set of exciton resonances in optical spectra~\cite{Gammon-Science1996}. The diffusion of atoms through the QW interface (segregation) during the growth process gives rise to smoothing and an asymmetry of potential profiles for carriers~\cite{Moison-PRB1989, Muraki-APL1992}. The specially designed asymmetric QWs are also extensively studied in view of their interesting properties, e.g., large electron spin-orbit splitting~\cite{Bassani-PRB1997, Grundler-PRL2000}, enhanced optical nonlinearity~\cite{Rosencher-PRB1991, Atanasov-PRB1994, Sun-PRB2013} and coupling with terahertz radiation~\cite{Bedoya-PRB2005, Liberato-PRB2013}. 

Exciton energy and wave function are sensitive to the potential profile. However the direct experimental observation of effects of the profile peculiarities on the exciton properties is difficult for several reasons. The exciton energy shift relative to the theoretically predicted value may be caused, apart from the modification of potential profile, by uncertainties in the QW width and in the composition of solid solution in the layers within the heterostructure. Various imperfections of the heterostructure like point defects, dislocations, etc., may broaden the exciton ensemble and complicate the study of exciton energies. The exciton wave function determines the exciton-light coupling, in particular, the oscillator strength and the radiative decay rate, which can be studied by reflectance spectroscopy and in the time-resolved experiments~\cite{Poltavtsev-SSC2013, Trifonov-PRB2015, Khramtsov-PhysE2016}. These are the integral characteristics, which are insensitive to peculiarities of potential profile for each particular exciton transition. At the same time, these difficulties can be overcome when several exciton transitions in the same QW are experimentally studied and analyzed.  

In this paper we experimentally study and theoretically analyze reflectance spectra of heterostructures with asymmetric InGaAs/GaAs QWs. We demonstrate that the simultaneous analysis of several exciton resonances in the spectra of relatively wide QWs allows one to obtain valuable information about the potential profile for the excitons. We have developed an approach for the direct numerical solution of Schr\"{o}dinger equation for an exciton in a QW with an arbitrary potential profile. The Coulomb electron-hole interaction in the exciton is exactly included into the numerical approach with no approximations. Calculations of the exciton energies and wave functions allowed us to accurately model reflectance spectra with the use of only two fitting parameters describing the maximal indium content in the QW and the characteristic length of the indium diffusion. Both these parameters cannot be controlled  during the growth of heterostructires with the high enough accuracy required for the modeling. The obtained agreement allowed us to reliable model the potential profile for excitons in the heterustructures under study.

The paper has the following structure. First we present experimental details and obtained reflectance spectra. A generalization of phenomenological theory of exciton-light coupling for the case of several exciton resonances in a QW with arbitrary potential profile is given in the next section. Then we compare experimental results and microscopic modeling of the exciton states. Finally we sum up major results in the Conclusion.

\section{experiment}
\label{experiment}

 We have experimentally studied reflectance spectra of several InGaAs/GaAs heterostructures grown by the molecular beam epitaxy (MBE). Two of the structures with the smallest inhomogeneous broadening of exciton resonances have been selected for detailed investigation. The first one (S1) contains a nominally square QW of 95-nm width with small indium content of about 2\% grown between the GaAs barrier layer. The second structure, S2, is the specially designed asymmetric InGaAs/GaAs QW with one vertical potential wall and other sloping potential wall. In all the structures, effects of the QW asymmetry are found in the comparative study of exciton resonances observed in reflectance spectra. The spectra have been measured using a femtosecond Ti:Sapphire laser or a halogen lamp as a light source. In the latter case, the light was focused onto a 50-$\mu$m pin-hole and then refocused onto the sample. The light was directed to the samples at small angle close to the normal incidence. The light spots on the samples were of about 100~$\mu$m in both cases. The samples were held in a vacuum chamber of a closed cycle cryostat at $T = 4$~K. The reflected light was dispersed in a 0.55~m spectrometer with the 1800 grooves/mm grating and detected by a nitrogen cooled CCD-matrix. The spectral resolution was of about 30~$\mu$eV. To obtain the absolute value of reflectivity, the reflectance coefficient was carefully measured for a single wavelength nearby an exciton resonance using a beam of a continuous wave Ti:sapphire laser focused at the same spot on the sample.

\begin{figure}
\includegraphics[scale=1]{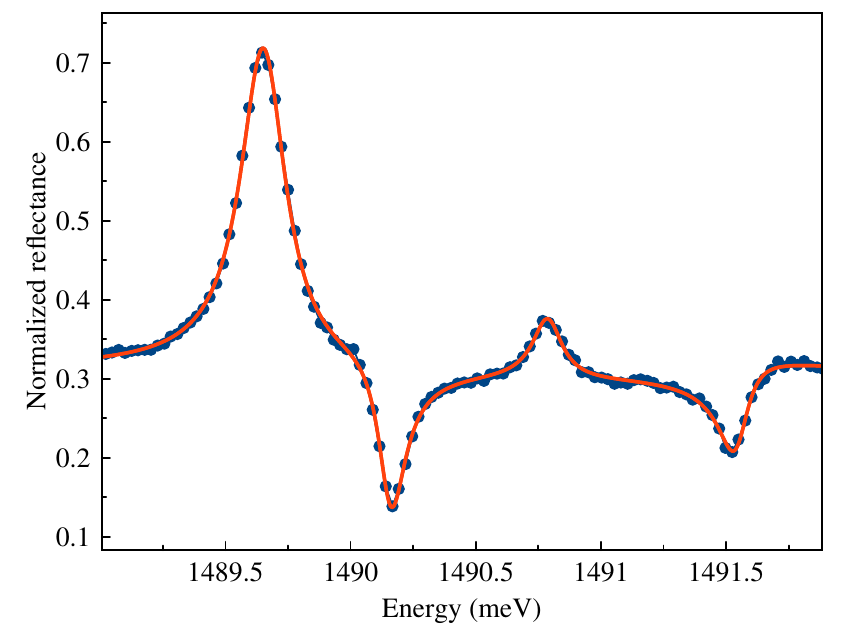}
\caption{Reflectance spectrum of InGaAs/GaAs heterostructure with the 95-nm square QW. The red line corresponds to the fit by a phenomenological model with the four free parameters for each resonance.} \label{Flo:figure1}
\end{figure}

\begin{figure}
\includegraphics[scale=1]{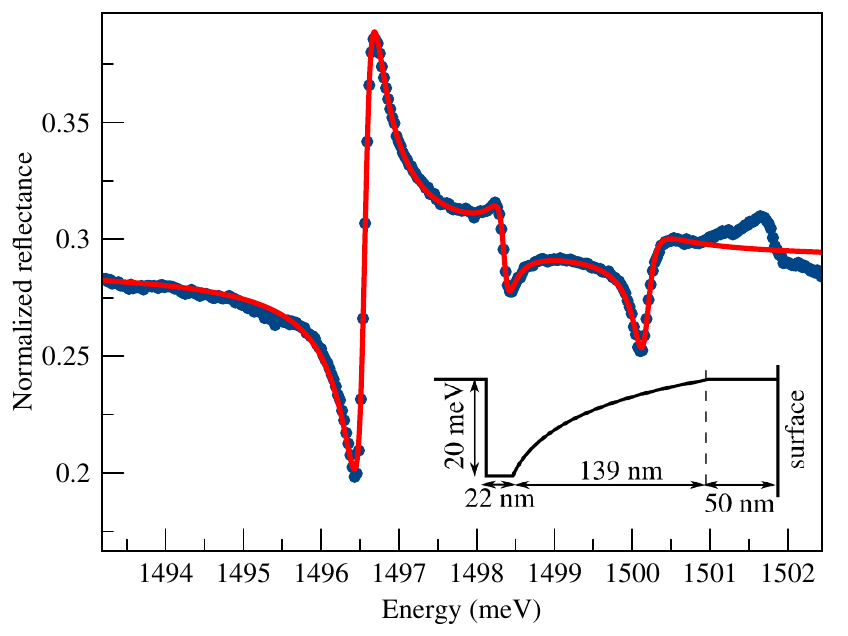}
\caption{Reflectance spectrum of heterostructure with the triangle-like QW (points) and its fit (red line). Inset represents the QW potential profile for excitons.}
\label{Flo:figure2}
\end{figure}

The reflectance spectra for samples S1 and S2 are shown in Figs.~\ref{Flo:figure1} and \ref{Flo:figure2}, respectively. For sample S1, four exciton resonances are clearly seen, two of them as the peaks and two others as the dips. The resonances correspond to optical transitions to the quantum confined excition states in the 95-nm QW. As we discuss in the next section, the different profile of the resonances (peak or dip) is due to the different phase shift in the light wave reflected form the QW at different exciton frequencies.  
 
In the spectrum of sample S2, three similar exciton transitions are clearly seen as the resonances. The profiles of these resonances are more complex and contain valuable information about the phase shifts of light wave in this highly asymmetric QW. 

The simplified potential profile for excitons in this QW is shown as an inset in Fig.~2. The vertical wall of the QW was formed during the growth by rapid opening of the In effusion cell held at $T_0 = 754$ $^{\circ}$C. The sloping wall was formed by slow cooling the cell down to $T_{\rm{fin}} = 669$ $^{\circ}$C with the constant rate so that the In flux was exponentially decreasing. Correspondingly, the indium concentration has been varied according to a phenomenological formula:
\begin{equation}
x_{\rm{In}}(z) = x_{max} \exp[(T_{\rm{In}}(z) - 755.5)/38.54]. 
\label{In-content}
\end{equation}
Here $T_{\rm{In}}(z)$ is the cell temperature in the Celsius scale, which was linearly decreased in time  and, correspondingly, along the growth axis $z$, $T_{\rm{In}}(z) = T_0 - a z$, with rate $a  = 0.6115$~$^{\circ}$C/nm. A relatively hight substrate temperature, $T = 555$ $^{\circ}$C, results in the noticeable segregation of In, which further complicates the potential profile. It will be discussed in Sect.~\ref{numerical}.  

\section{Phenomenological analysis}
\label{phenomen}

A phenomenological model of resonant reflectance related to a single exciton transition in the symmetric QW is well developed~\cite{Andreani-SSC1991, Ivchenko-FTT1991}. Here we follow the basic theory described in Ref.~\cite{Ivchenko-book}. The reflectance is determined by the interference of the light waves reflected from the sample surface and the QW layer:
\begin{equation}
	R=\left|\frac{r_{01}+r_{QW}e^{i\varphi}}{1+r_{01}r_{QW}e^{i\varphi}}\right|^2
	\label{eqn:1}
\end{equation}
 Here $r_{01}$ and $r_{QW}$ are the amplitude reflectance coefficients from the sample surface  and from the QW, respectively. Phase shift $\varphi$ is determined by the distance from the surface to the QW center: 
 \begin{equation}
 \varphi = 4\pi (L_b + L_{QW}/2) \sqrt{\varepsilon_b}/\lambda,
 \label{eqn:varphi}
 \end{equation}
 where $L_b$ is the thickness of the top barrier layer, $L_{QW}$ is the thickness of QW layer, $\varepsilon_b$ is the dielectric constant of the layers, and $\lambda$ is the light wavelength. 

An analytical solution of the Maxwell equation for the light interaction with an exciton in a QW gives rise to the following equation for the amplitude reflectance~\cite{Ivchenko-book}:
\begin{equation}
	r_{QW}=\frac{i\varGamma_{0}}{\tilde\omega_{0}-\omega-i(\varGamma+\varGamma_{0})}
	\label{eqn:2}
\end{equation}
with $\varGamma_{0}$ and $\tilde\omega_{0}$ defined as:
\begin{equation}
\varGamma_0=\frac{\pi}{2}q\omega_{LT}a_B^3 \left[\int{\varPhi(z)\cos(q z)\,dz}\right]^2,
\label{eqn:3}
\end{equation}
\begin{equation}
\tilde\omega_0=\omega_0+\frac{\pi}{2}q\omega_{LT}a_B^3 \int\int\varPhi(z)\varPhi(z')\sin(q|z-z'|) dz dz'.
\label{eqn:omega0}
\end{equation}
Here $\omega_0$ is the exciton resonance frequency, $\varGamma_0$ is radiative decay rate and $\varGamma$ is the phenomenologically introduced nonradiative broadening~\cite{comment-broadening}. Function $\varPhi(z)$ is the amplitude of exciton wave function when the coordinates of an electron and a hole in the exciton coincide. Quantity $q$ is the wave vector of light, $\omega_{LT}$ is the 3D-exciton longitudinal-transverse splitting, and $a_B$ is the exciton Bohr radius. 

Equations (\ref{eqn:2}-\ref{eqn:omega0}) are applicable for a single exciton transition in a symmetric QW. They should be generalized for our case of multiple exciton transitions in asymmetric QWs. For the quantum-confined exciton states separated by energy distance, $\Delta E > \varGamma_0 + \varGamma$, a sum over the exciton resonances should be considered~\cite{Trifonov-PRB2015, comment-1}:
\begin{equation}
	r_{QW}=\sum_{n=1}^{n_{max}}{\frac{i \varGamma_{0n}e^{i\phi_n}}{\tilde\omega_{0n}-\omega-i(\varGamma_n+\varGamma_{0n})}}
	\label{eqn:rQW}
\end{equation}

An asymmetry of the QW potential requires an additional phase shift $\phi_n$ in the numerator of Eq.~(\ref{eqn:rQW}) (see Appendix~\ref{maintheory} for details) and  a generalization of Eq.~(\ref{eqn:3}): 
\begin{eqnarray}
\varGamma_{0n} &=& \frac{\pi}{2}q\omega_{LT}a_B^3 \left(\left[\int{\varPhi_n(z)\sin(q z)\,dz}\right]^2\right. \nonumber \\ 
&+& \left.\left[\int{\varPhi_n(z)\cos(q z)\,dz}\right]^2\right).
\label{eqn:Gamma0}
\end{eqnarray}
As seen the two exciton-light overlapping integrals appear in the case of asymmetric potential. When the QW potential is symmetric, only one integral containing $\cos(qz)$ for exciton states $n = 1, 3, \ldots$ or  $\sin(qz)$ for exciton states $n = 2, 4, \ldots$ is non-zero.

Phase shift $\phi_n$ is determined by ratio of the integrals:
\begin{eqnarray}
\label{eqn:phases}
\tan{\frac{\phi_n}{2}} &=& \frac{\int{\varPhi_n(z)\sin{(q z)}\,dz}}{\int{\varPhi_n(z)\cos{(q z)}\,dz}}.
\end{eqnarray}
As seen this expression also includes both $\cos{(q z)}$  and $\sin{(q z)}$. Phases $\phi_n$ can be used as a measure of the asymmetry of the potential profile. In the symmetric QW, they all should be zero for exciton states $n = 1, 3, \ldots$ and $\pi$ for states $n = 2, 4, \ldots$ However in the experiment they can be determined up to a constant phase shift $\varphi$ defined by Eq.~(\ref{eqn:varphi}). Problem is that the QW thickness $L_{QW}$ cannot be accurately defined for the asymmetric QW. We, therefore, consider phases 
\begin{equation}
\tilde{\phi}_n = \phi_n + \varphi,
\label{eqn:tildephi}
\end{equation}
which are directly determined from the experimental spectra fitting them by Eqs.~(\ref{eqn:1}) and (\ref{eqn:rQW}). 

We have fitted several exciton resonances observed in the reflectance spectra of the heterostructures studied. Results are shown in Figs.~\ref{Flo:figure1} and \ref{Flo:figure2}. Respective fitting parameters are listed in Tabs.~\ref{P554_table} and \ref{T694_table}. As seen from the figures, the fit by Eqns.~(\ref{eqn:1}), and (\ref{eqn:rQW}) allow us accurately reproduce the spectra for both samples in the spectral range of several excitonic resonances. We did not fit the fourth exciton resonance for the triangle QW (see the spectral peculiarity in the range 1501 -- 1502 meV in Fig.~\ref{Flo:figure2}) because it is superimposed on the 2s-exciton transition not analyzed in present work. The obtained results support the generalization of the phenomenological model suggested. We also may conclude that the parameters obtained in the fitting are reliably determined. 

As seen from Tab.~\ref{P554_table} for the square QW, phases $\tilde{\phi}_n$ for states $n = 1, 3$ are close to each other and small while for states $n = 2, 4$  they are close to $\pi$. This is an indication that the asymmetry of this QW in not large. In the case of the triangular QW (see Tab.~\ref{T694_table}), the phase difference is much larger that clearly shows sensitivity of the phases to the QW asymmetry. 
The quantitative analysis of the phases will be done in the next section after the microscopic modeling of the exciton states.
 
\begin{table}[h]
\begin{ruledtabular}
\caption{Fitting parameters extracted from the experiment for the square QW versus those obtained in the microscopic modeling. Parameters of the modeling: $L_b=70$~nm, $L_{QW}=95$~nm, $x_{max} = 1.67\,$\%, $\lambda = 3.75$~nm. The common phase shift $2\pi$ is subtracted from phases $\tilde{\phi}_n$. \label{tbl:1}}
\label{P554_table}
\begin{tabular}{ c c c c c c }
& & X1 & X2 & X3 & X4 \\
\hline
$\hbar \tilde{\omega}_{0n}$ \text{(meV)} & Exp. & 1489.65 & 1490.16 & 1490.79 & 1491.55 \\
& Comp. & 1489.69 & 1490.15 & 1490.80& 1491.61\\
\hline
$\hbar\varGamma_{0n}$ \text{($\mu$eV)} & Exp. & 47.2 & 19.1 & 6.9 & 11.1 \\
& Comp. & 50.8 & 18.8 & 4.3 & 8.2\\
\hline
$\hbar\varGamma_n$ ($\mu$eV) & Exp. & 37.7 & 59 & 61 & 67 \\
\hline
$\tilde{\phi}_n$ (rad) & Exp. & $0.16$ & $3.10$ & $0.14$ & $3.70$ \\
& Comp. & 0.16 & 3.29 & 0.23 & 3.29\\
\end{tabular}
\end{ruledtabular}
\end{table}

\begin{table}[h]
\begin{ruledtabular}
\caption{Fitting parameters extracted from the experiment for the triangular QW versus those obtained in microscopic modeling.  In the microscopic modeling we used: $L_b=50$~nm, $L_{1}=22$~nm, $L_2 = 139$~nm, $x_{max} = 1.32\,$\%, $\lambda = 4.5$~nm.  \label{tbl:2}}
\label{T694_table}
\begin{tabular}{ c c c c c c }
& & X1 & X2 & X3 \\
\hline
$\hbar \tilde{\omega}_{0n}$ \text{(meV)} & Exp. & 1496.581 & 1498.355 & 1500.170  \\
& Comp. & 1496.61 & 1498.16 & 1499.55\\
\hline
$\hbar\varGamma_{0n}$ \text{($\mu$eV)} & Exp. & 31.5 & 4.0 & 8.1 \\
& Comp. & 36.3 & 7.1 & 9.8\\
\hline
$\hbar\varGamma_n$ ($\mu$eV) & Exp. & 99 & 84 & 124 \\
\hline
$\tilde{\phi}_n$ (rad) & Exp. & $4.61$ & $8.15$ & $3.84$ \\
& Comp. & 4.61 & 8.2 & 3.78\\
\end{tabular}
\end{ruledtabular}

\end{table}

\section{Microscopic modeling}
\label{numerical}

We consider the problem of an exciton in a QW described by the stationary Schr\"{o}dinger equation for two particles (electron and hole) with Coulomb interaction. The wave function of the exciton can be expressed in planar heterostructures as:
\begin{equation}
	\psi(X,Y,z_e,z_h,\rho,\varphi)=e^{i K_X X + i K_Y Y}\psi(z_e,z_h,\rho)e^{i k_\varphi \varphi}
	\label{wf}
\end{equation} 
with $X$ and $Y$ as the center-of-mass coordinates in the XY plane, $z_e$ and $z_h$ as the electron and hole coordinates in the growth direction, respectively. Quantities $\rho$ and $\varphi$ are the polar coordinates of relative electron-hole motion defined in the plane perpendicular to the growth axis. Quantity $k_\varphi$ is the $z$-projection of the exciton angular momentum. We suggest it to be zero for the optically observable exciton states ($s$-like states).

Function $\psi(z_e,z_h,\rho)$ is the eigenfunction of operator:
\begin{equation}
	\hat{H}=\hat{K}+V_e f(z_e)+V_h f(z_h)-\frac{e^2}{\varepsilon \sqrt{\rho^2+(z_e-z_h)^2}}.
\label{numericalop}
\end{equation}
Here $\varepsilon = 12.56$ is the dielectric constant for GaAs, $f(z)$ is the QW potential function, $V_h$ and $V_e$ are the valence and conduction band offsets, respectively, at the point where $f(z)=1$. We assume that $V_e=2V_h$ for all of the In concentrations considered. To calculate the potential well for excitons we use a phenomenological dependence of the band gap for solid solution In$_x$Ga$_{1-x}$As on the indium concentration $x$~\cite{materials}:
\begin{equation}
	E_g(x) = E_g(\mbox{InAs}) x + E_g(\mbox{GaAs}) (1-x) - 0.477 x (1-x),
\end{equation}
where $E_g(\mbox{GaAs}) = 1.519$~eV and $E_g(\mbox{InAs}) = 0.417$~eV are the band gaps for GaAs and InAs, respectively. The bottom of potential well for excitons is determined as 
\begin{equation}
	V(x) \equiv V_e + V_h = E_g(\mbox{GaAs}) - E_g(x).
\end{equation}
The profile of QW potential is determined by function $f(z)$, which will be discussed below.

Operator $\hat{K}$  in Eq.~(\ref{numericalop}) consists of three terms:
\begin{equation}
	\hat{K}=-\frac{\hbar^2}{2 m_{e}}\frac{\partial^2}{\partial z_e^2}-\frac{\hbar^2}{2 m_{z_h}}\frac{\partial^2}{\partial z_h^2}-\frac{\hbar^2}{2 \mu_{xy}}\Delta_\rho.
\end{equation}
Here $\Delta_\rho$ is the Laplacian in polar coordinates for the case $k_{\varphi}=0$: 
\begin{equation}
	\Delta_\rho=\frac{1}{\rho}\frac{\partial}{\partial\rho}\left(\rho\frac{\partial}{\partial\rho}\right)
\end{equation}

In the GaAs-based heterostructures, the conduction band can be considered as an isotropic one with an effective electron mass $m_{e}=0.0665~m_0$. The valence band is twice degenerate, consists of heavy-hole and light-hole subbands, and is described in terms of the Luttinger Hamiltonian~\cite{Luttinger}. The effect of quantum confinement breaks the valence band degeneracy, which results in the anisotropic heavy-hole and light-hole masses. The InGaAs/GaAs QWs are also affected by a strain due to the mismatch of lattice constants of InAs and GaAs crystals, which induces the heavy-hole-light-hole splitting up to 10~meV for the 2\% In concentration in the QWs~\cite{Van_De_Walle}. This splitting effectively reduces the heavy-hole-light-hole interaction, therefore we can introduce the heavy hole masses in growth direction and in xy-plane, respectively: $m_{z_h}=m_0/(\gamma_1-2\gamma_2)$ and $m_{xy_h}=m_0/(\gamma_1+\gamma_2)$. 
The reduced exciton mass in xy-plane $\mu_{xy} = m_{xy_h} m_e/(m_{xy_h}+m_e)$. 
Due to the low indium concentration, we use the GaAs Luttinger parameters suggested by Vurgaftman~\textit{et al.}~\cite{materials}: $\gamma_1=6.95$, $\gamma_2=2.06$.

The QW potential profile defined by $f(z)$ function is known to be significantly modified by the segregation effect in the InGaAs/GaAs QWs~\cite{Muraki-APL1992, Martini-APL2002, Drozdov-FTP2003, Martini-JVST2010}. The indium atoms are more mobile during the growth process as compared to the gallium atoms. This results in a diffusion of the indium atoms from lower layers of growing structure to the higher ones. Related modification of the QW potential can be phenomenologically well described using only a single parameter, $\lambda_D$~\cite{Muraki-APL1992}. This parameter is the indium diffusion length, which characterizes the exponentially decaying indium concentration after deposition of the $\delta$-layer of In: $x(z) = x(0) \exp(-z/\lambda_D)$. It depends on the growth conditions, in particular, on the substrate temperature and varies in the range $\lambda_D = 1.5 \ldots 4.5$~nm for temperatures $T = 500 \ldots 550$~$^\circ$C. We use this parameter as the free one in the modeling of exciton spectra.

In the framework of the phenomenological model, the segregation process can be described by a simple rate equation for In concentration, $x(z)$, as a function of coordinate $z$ along the growth direction:
\begin{equation}
\frac{dx(z)}{dz} = -\frac{x(z)}{\lambda_D} + \frac{F_{\rm{In}}(z)}{\lambda_D}.
\label{eqn:segregation}
\end{equation}
Here $F_{\rm{In}}(z)$ is the In flux with taking into account the sticking coefficient. The flux is varied during the growth process in a general case.

The general solution of Eq.~(\ref{eqn:segregation}) is:
\begin{equation}
x(z) = e^{-z/\lambda_D} \int_{z_0}^z e^{z'/\lambda_D}\frac{F_{\rm{In}}(z')}{\lambda_D}\,dz'.
\label{eqn:segr_solution}
\end{equation}
Here $z_0$ is the coordinate where the indium containing layers start to grow.

In the particular case of square QW, the indium flux $F_{\rm{In}}(z) = F_{\rm{In}}^{0}$ within the QW and zero outside it. The integration in Eq.~(\ref{eqn:segr_solution}) gives rise to solution:
\begin{eqnarray}
\label{eqn:segr_squareQW}
x(z) &=& F_{\rm{In}}^0 \left(1 - e^{-z/\lambda_D}\right), \quad 0 < z < L_{QW}, \\
x(z) &=& F_{\rm{In}}^0 \left(1 - e^{-L_{QW}/\lambda_D}\right) e^{-(z-L_{QW})/\lambda_D}, \quad z > L_{QW}. \nonumber
\end{eqnarray}
Here $L_{QW}$ is the nominal QW width. This solution coincides with that presented in Ref.~\cite{Muraki-APL1992}.

\begin{figure}
\includegraphics[scale=1]{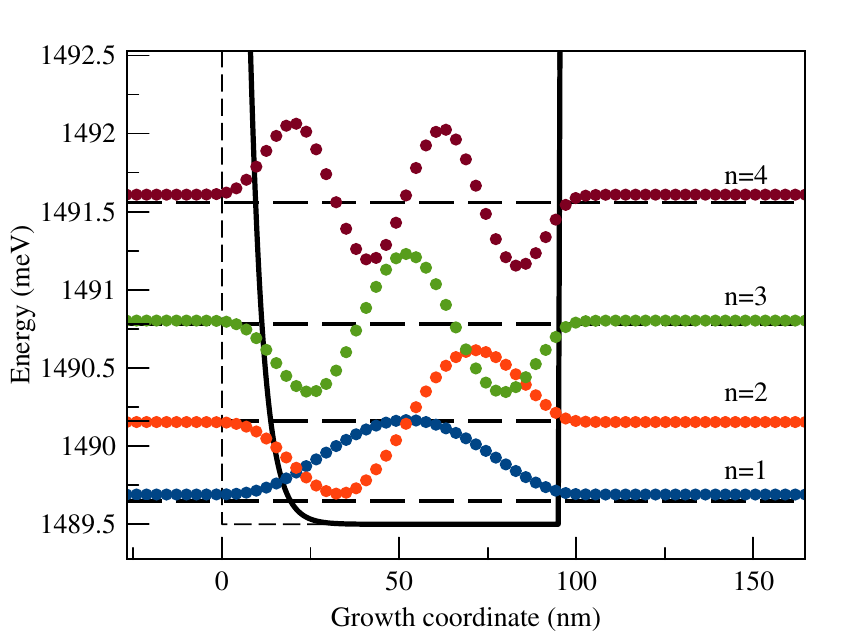}
\caption{Segregated QW potential for excitons (black solid line) and functions $\varPhi(z)$ (colored dots) for the first four excitonic states. The thin dashed line shows an initial profile of the square QW. The dashed horizontal lines correspond to the exciton energies extracted from the reflectance spectrum shown in figure~\ref{Flo:figure1}.}
\label{Flo:figure3}
\end{figure}

The potential function $f(z)$ [see Eq.~(\ref{numericalop})] is expressed via $x(z)$:
\begin{equation}
f(z) = 1 - \frac{x(z)}{x_{max}},
\label{eqn:segr_f}
\end{equation}
 where $x_{max}$ is the In concentration in the bottom of potential well. The potential profile for excitons in square QW under study modeled by this way is shown in  Fig.~\ref{Flo:figure3}.

\begin{figure}
\includegraphics[scale=1]{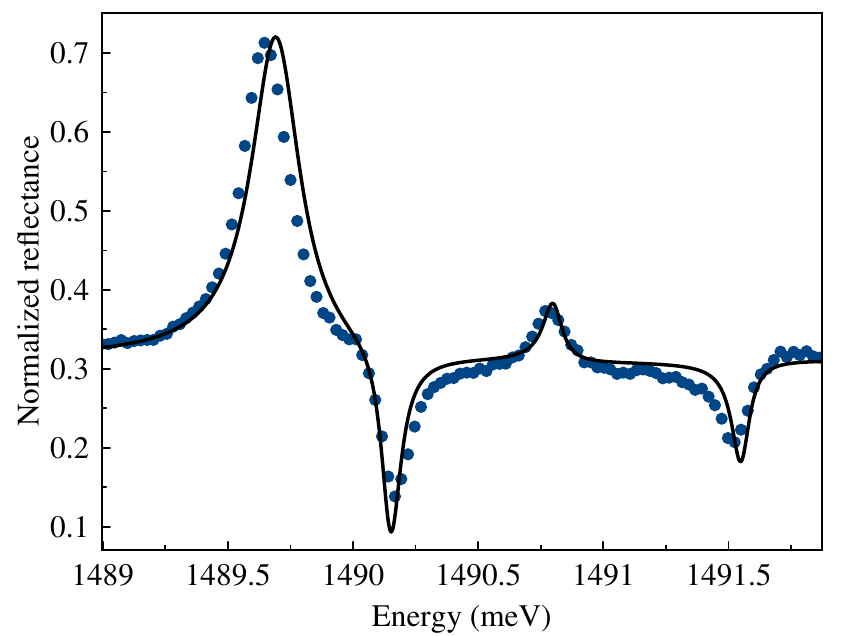}
\caption{The comparison of theoretically modeled reflectance spectrum of InGaAs/GaAs heterostructure with the the square QW (solid line) and that measured experimentally (line with points). The theoretical spectrum is calculated using Eq.(\ref{eqn:1}) and parameters for exciton resonances obtained in microscopic calculation (see table~\ref{tbl:1}). Parameters of non-radiative broadening, $\hbar\Gamma_n$, have been taken 40\,$\mu$eV for all the resonances.}
\label{Flo:figure4}
\end{figure}

Using the segregated function $f(z)$, we find several lowest eigenstates of the eigenproblem with operator~(\ref{numericalop}). We represent the operator as a matrix composed according to the 3-point finite difference representation of differential operators. Computation area is $200 \times 200 \times 400$~nm$^3$ with $70 \times 70 \times 200$~points along the $z_e$, $z_h$, and $\rho$ coordinates, respectively. An Arnoldi algorithm realization in the ARPACK library was used to obtain the eigenstates. Further technical details of the numerical calculations can be found in Ref.~\cite{Khramtsov-JAP2016}.

For the square QW potential smoothed by the segregation, we varied the depth of the QW potential, $V = V_e + V_h$, see Eq.~(\ref{numericalop}), and the In diffusion length $\lambda_D$ to obtain the best correspondence of calculated energies of the quantum confined states with those found experimentally. We should note that the variation of parameter $V$ mainly gives rise to the common energy shift of exciton levels with small change of energy gaps between them. This is due to the relatively deep potential well, $V \approx 26$~meV, relative to the energy range for exciton states under study, see Fig.~\ref{Flo:figure3}. For the sake of graphical presentation we shifted the QW potentials in Figs.~\ref{Flo:figure3} and~\ref{Flo:figure4} down by the $4.2\,$meV value (the bulk exciton binding energy).

The energy difference between the neighboring exciton levels, $\Delta E_{nn'} = \hbar\tilde{\omega}_{n'} - \hbar\tilde{\omega}_{n}$, in particular, the difference $\Delta E_{12}$, is very sensitive to parameter $\lambda_D$, because of sensitivity of curvature of the potential near the bottom, see Fig.~\ref{Flo:figure3}. When one neglects the segregation, the energy difference $\Delta E_{12} = 0.41$~meV that is noticeable smaller than the difference of 0.51~meV obtained from the experiment (see Tab.~\ref{P554_table}). These properties of exciton spectrum allow one to obtain  parameters $V$ and $\lambda_D$ independently. 

The calculated energies are shown in figure~\ref{Flo:figure3} and also present in Tab.~\ref{P554_table}. As seen, the calculated and measured energies coincide with the accuracy in several tens of $\mu$eV. We would like to stress that this correspondence has been achieved by the variation of only two fitting parameters.  

The figure also shows functions $\varPhi_n(z)$ obtained in the numerical computations. They are shifted vertically according to the calculated energies. We should note one more effect of segregation, namely, the stronger penetration of functions $\varPhi_n(z)$ to the left barrier compared to the right one.

The obtained functions $\varPhi_n(z)$  allowed us to calculate radiative broadenings, $\hbar\varGamma_{0n}$, using Eq.~(\ref{eqn:Gamma0}). They are compared with the experimentally obtained data in table~\ref{P554_table}. As seen the calculated radiative broadenings  are in good agreement with the experimental data. 

Phases $\phi_n$ have been calculated by the use of Eqs.~(\ref{eqn:phases}). A constant phase $\varphi$ has been added to phases $\phi_n$ to take into account the phase shift during propagation of light wave from the sample surface to the QW and back, see Eq.~(\ref{eqn:varphi}). We have chosen phase $\varphi$ so that phase $\tilde{\phi}_1$ is equal to the experimentally obtained value.  
The calculated phases are close to those obtained from the experiment for all four resonances in this structure. There is some discrepancy with the phase for the fourth resonance obtained from the experiment. The possible reason for that is a complexity of experimental spectrum in the range of this resonance, which can be superimposed on the transition to the 2s exciton state originated from the lowest quantum confined exciton state.

The calculated parameters $\hbar\varGamma_{0n}$ and $\phi_n$ have been used to simulate, exploiting Eq.~(\ref{eqn:1}), the reflectance spectrum for structure with the square QW. The only parameters additionally taken are: the nonradiative broadening $\hbar\varGamma_{n} = 40$~$\mu$eV for all the resonances~(compare with Tab.~\ref{P554_table}), the effective top barrier layer $L_b = 70$~nm defining phase $\varphi$ [see Eq.~(\ref{eqn:varphi})], and the dielectric constant $\varepsilon_b = 12.56$ defining the reflectance beyond the exciton resonances. Results of the theoretical modeling are compared with the experiment in Fig.~\ref{Flo:figure4}. As seen, the modeling accurately reproduces all the features of the experimental spectrum.

\begin{figure}
\includegraphics[scale=1]{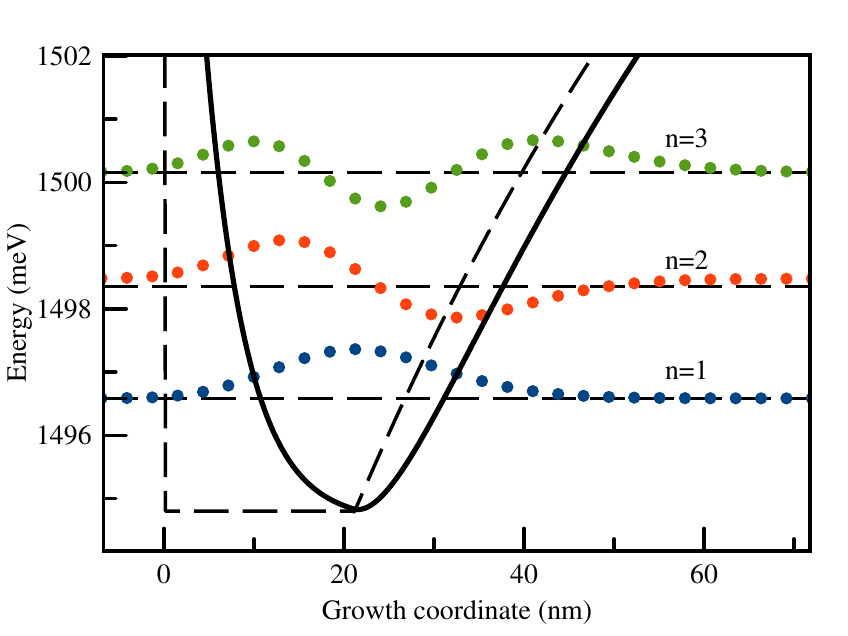}
\caption{Segregated asymmetric QW potential for excitons and $\varPhi(z)$ wave functions of first three states. Dashed horizontal lines correspond to energies extracted from spectrum on figure~\ref{Flo:figure2}.
\label{Flo:figure5}
}
\end{figure}

\begin{figure}
\includegraphics[scale=1]{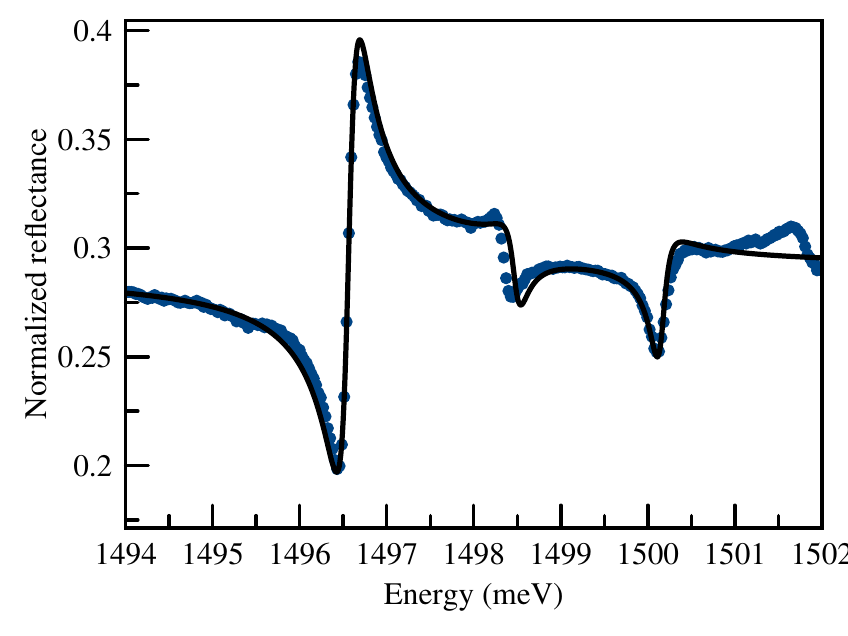}
\caption{Reflectance spectrum of heterostructure with the asymmetric QW. Black line is the spectrum obtained using Eq.(\ref{eqn:1}). Parameters for each exciton resonance were obtained in microscopic calculation (see table~\ref{tbl:2}). As $\hbar\Gamma_n$ parameter we used 100\,$\mu$eV for all resonances.}
\label{Flo:figure6}
\end{figure}

Similar microscopic modeling has been done for sample S2 with the triangle-like QW.  
Segregation model in this case can be easily calculated using general solution~(\ref{eqn:segr_solution}).
The obtained potential profile is shown in Fig.~\ref{Flo:figure5}. This figure also demonstrate the  calculated energies as well as functions $\varPhi_n(z)$ for the quantum-confined excitonic states experimentally studied. The discrepancy of the calculated energies with the measured ones is also negligibly small (see Tab.~\ref{T694_table}) that indicates that the potential profile of the QW is correctly modeled. The microscopic modeling also allowed us to calculate the radiative broadenings and phase shifts for the exciton transitions. They are compared with the experiment in  table~\ref{T694_table}. As seen good agreement is observed for the data. 

 Using the calculated quantities, we have simulated the reflectance spectrum for this sample. It is shown in Fig.~\ref{Flo:figure6}. Good correspondence of the calculated and experimental spectra is observed that supports the proposed model of potential profile.  

\section{Conclusion}

We have experimentally studied and theoretically modeled several exciton resonances in reflectance spectra of three samples with asymmetric InGaAs/GaAs QWs. The first sample (S1) contains a nominally symmetric square QW of 95-nm width. The segregation of indium during the MBE growth of the structure results in an asymmetry of the QW potential profile. This asymmetry reveals itself in a change of energy gaps between the neighboring quantum-confined excitonc levels. No valuable change of phases is observed for this sample.

The study of sample with the triangle-like QW (S2) show that both the energy position of exciton resonances and their phases are very sensitive to the potential profile. This sensitivity allows one to reliably model the potential profile exploiting only a few free parameters. The important parameters are the indium content at the bottom of QWs, $x_{max}$, and the segregation diffusion length, $\lambda_D$. The analysis performed shows that the exciton resonances are highly sensitive to these parameters so that the accuracy of their determination by other methods are not sufficient for the theoretical modeling of the potential well.

The simulated reflectance spectra are in the good agreement with the experimentally observed ones. This means that all the valuable processes are included in the modeling. In fact, the key process in the structures studied is the exciton-light coupling determining radiative broadening of exciton resonances, $\hbar\varGamma_{0n}$, and the phase shifts, $\phi_n$, of light wave during the reflection from the QW. The nonradiative broadening, $\hbar\varGamma_{n}$, cannot be modeled in the framework of the proposed approach. However, the relatively small difference of their values for different exciton states and a slow monotonic rise with number of the exciton state indicates that most probable origin of the broadening is the exciton-phonon scattering. Further studies are needed to model this process.

 We also generalized the phenomenological model for description of multiple exciton resonances in the spectra of QWs with arbitrary potential profile. This generalization has been verified by the analysis of the reflectance spectra and allowed us to perfectly fit the spectra and to obtain reliable values all the quantities describing the resonances. 
 
 Finally we should stress that the study of several exciton resonances allows one to reliably determine the potential profile of the QW localizing the excitons. Both the exciton energies and the light phases contain valuable information about the profile. 
 
 \section*{acknowledgments}
 The authors thank I.~Ya.~Gerlovin and M.~V.~Durnev for fruitful discussions. Financial support from SPbU (grant No.~11.38.213.2014)  and RFBR (grant No.~16-02-00245) is acknowledged.
 P.S.G. and A.V.K. thank the financial support
of RFBR (grant No. 15-52-12019) and DFG in the frame of Project ICRC TRR 160. The authors also thank the SPbU Resource Center
``Nanophotonics'' (www.photon.spbu.ru) for the sample studied in present
work. 

\appendix

\section{Exciton-induced resonant reflectance}
\label{maintheory}

The amplitude reflection coefficient for a QW, $r_{QW}$, arises from the solution of wave equation for electric field of light, $E$, in heterostructure:
\begin{equation}
\frac{d^2E}{dz^2}=-\left(\frac{\omega}{c}\right)^2 \left[\varepsilon_b E+4\pi P_{exc}(z) \right],
\label{f.Elect}
\end{equation}
where $P_{exc}(z)$ is a nonlocal dielectric polarization of exciton given by equations~\cite{Ivchenko-book}:
\begin{eqnarray}
\label{f.Pexc}
4\pi P_{exc}(z) &=& G(\omega) \varPhi(z)\int \varPhi^*(z') E(z') dz', \\ \nonumber
G(\omega) &=& \frac{\pi\varepsilon_b\omega_{LT} a_B^3}{\hbar(\omega_0-\omega-i\varGamma)}
\end{eqnarray}
Function $\varPhi(z) = \varPhi^*(z)$ is defined as
\begin{equation}
\label{f.varphi}
\varPhi(z)=\varphi(z,z,0),
\end{equation}
where $\varphi(z_e,z_h,\rho)$ is the part of exciton wave function~(\ref{wf}), which satisfy a Schr{\"o}dinger equation with Hamiltonian~(\ref{numericalop}). 

The solution of equation (\ref{f.Elect}) with function $P_{exc}(z)$ is  given by formula~\cite{Ivchenko-book}:
\begin{equation}
\label{f.solvE}
\begin{split}
E(z)=& E_0e^{iqz}+ i\frac{q_0^2}{2q} G(\omega) \varLambda \int e^{iq|z-z'|}\varPhi(z') dz',
\end{split}
\end{equation}
where $q_0=\omega/c$ and $q=(\omega/c)\sqrt{\varepsilon_b}$ are the light wave vectors in vacuum and in the QW, respectively. The first term in this equation describes the light wave with amplitude of electric field $E_0$ incident on the QW. The second term is the secondary wave induced by the exciton in the QW. Quantity $\varLambda$ is:
\begin{equation}
\label{f.Lambda}
\varLambda=\frac{E_0 \int \varPhi(z) e^{iqz} dz}{1 - i q_0^2/(2q) G(\omega) \int \int e^{iq|z-z'|} \varPhi(z) \varPhi(z') dz dz'}.
\end{equation}

To contrast to Ref.~\cite{Ivchenko-book} we do not assume any parity of function $\varPhi(z)$ and, therefore, keep function $\exp(iqz)$ in the numerator of this expression rather than its even part, $\cos(qz)$, compare with Eq.~(3.17) in Ref.~\cite{Ivchenko-book}. Using definitions for $\tilde{\omega}_0$, Eq.~(\ref{eqn:omega0}), and $\varGamma_0$, Eq.~(\ref{eqn:Gamma0}), we obtain:
\begin{equation}
\label{f.LamdaFin}
\varLambda=E_0\frac{(\omega_0-\omega-i\varGamma)\int\varPhi(z) e^{iqz} dz}{\tilde{\omega}_0-\omega-i(\varGamma+\varGamma_0)}.
\end{equation}

The coefficient of amplitude reflectance from the QW is the ratio of the second and first terms in Eq.~(\ref{f.solvE}):
\begin{equation}
\label{f.rQWfin}
r_{QW}=\frac{i \frac{\pi}{2}q\omega_{LT}a_B^3 \left[\int\varPhi(z)e^{iqz}dz\right]^2}{\tilde{\omega}_0-\omega-i(\varGamma+\varGamma_0)},
\end{equation}
The numerator of this equation can be expressed as $i \varGamma_0 \exp(i\phi)$, where $\phi$ is defined by Eq.~(\ref{eqn:phases}).
For symmetric QWs, function $\varPhi(z)$ is even or odd depending on the number $n$ of the quantum-confined state. Correspondingly, phase $\phi = 0$ or $\pi$. 
Summing contributions (\ref{f.rQWfin}) to the reflection from several exciton resonances, we obtain final expression~(\ref{eqn:rQW}).

\end{document}